# Competitive Safety Analysis: Robust Decision-Making in Multi-Agent Systems

**Moshe Tennenholtz**                                           MOSHET@IE.TECHNION.AC.IL
*Faculty of Industrial Engineering and Management*
*Technion – Israel Institute of Technology*
*Haifa 32000, Israel*

## Abstract

Much work in AI deals with the selection of proper actions in a given (known or unknown) environment. However, the way to select a proper action when facing other agents is quite unclear. Most work in AI adopts classical game-theoretic equilibrium analysis to predict agent behavior in such settings. This approach however does not provide us with any guarantee for the agent. In this paper we introduce competitive safety analysis. This approach bridges the gap between the desired normative AI approach, where a strategy should be selected in order to guarantee a desired payoff, and equilibrium analysis. We show that a safety level strategy is able to guarantee the value obtained in a Nash equilibrium, in several classical computer science settings. Then, we discuss the concept of competitive safety strategies, and illustrate its use in a decentralized load balancing setting, typical to network problems. In particular, we show that when we have many agents, it is possible to guarantee an expected payoff which is a factor of 8/9 of the payoff obtained in a Nash equilibrium. Our discussion of competitive safety analysis for decentralized load balancing is further developed to deal with many communication links and arbitrary speeds. Finally, we discuss the extension of the above concepts to Bayesian games, and illustrate their use in a basic auctions setup.

## 1. Introduction

Deriving solution concepts for multi-agent encounters is a major challenge for researchers in various disciplines. The most famous and popular solution concept in the economics literature is the Nash equilibrium. Although Nash equilibrium and its extensions and modifications are powerful descriptive tools, and have been widely used in the AI literature (Rosenschein & Zlotkin, 1994; Kraus, 1997; Sandholm & Lesser, 1995), their appeal from a normative AI perspective is somewhat less satisfactory.[1] We wish to equip an agent with an action that guarantees some desired outcome, or expected utility, without relying on other agents' rationality.[2] This paper shows that, surprisingly, the desire for obtaining a guaranteed expected payoff, where this payoff is of the order of the value obtained in a

---

1. If we restrict ourselves to cases where there exists an equilibrium in dominant strategies, as is done in some of the CS literature (Nisan & Ronen, 1999), then the corresponding equilibrium is appealing from a normative perspective. However, such cases rarely exist.
2. Maximizing expected payoff when facing a set of possible environment behaviors is fundamental to AI. In particular, it is discussed in the context of game trees, in the context of planning with incomplete information, where we need to obtain a desired goal regardless of the initial configuration, as well as in the context of reinforcement learning, where we wish to maximize expected payoff when the actual model (selected from a set of possible models in adversarial way) is initially unknown. (Russell & Norvig, 1995).

Nash equilibrium, is achievable in various classical computer science settings. Our results are inspired by several interesting examples for counter-intuitive behaviors obtained by following Nash equilibria and other solution concepts (Roth, 1980; Aumann, 1985). One of the most interesting and challenging examples has been introduced by Aumann (Aumann, 1985). Aumann presented a 2-person 2-choice ($2 \times 2$) game $g$, where the safety-level (probabilistic maximin) strategy of the game is not a Nash equilibrium of it, but it does yield the expected payoff of a Nash equilibrium of $g$. This observation may have significant positive ramifications from an agent's design perspective. If a safety-level strategy of an agent guarantees an expected payoff that equals its expected payoff in a Nash equilibrium, then it can serve as a desirable robust protocol for the agent! Given the above, we are interested in whether an optimal safety level strategy leads to an expected payoff similar to the one obtained in a Nash equilibrium of simple games that represent basic variants of classical computer science problems. As we show, this is indeed the case for $2 \times 2$ games capturing simple variants of the classical load balancing and leader election problems. A more general question refers to more general $2 \times 2$ games. We show that if the safety-level strategy is a (strictly) mixed one, then its expected payoff is identical to the expected payoff obtained in a Nash equilibrium in any generic non-reducible $2 \times 2$ game. We also show that this is no longer necessarily the case if we have a pure safety-level strategy. In addition, we consider general 2-person set-theoretic games (which naturally extend $2 \times 2$ leader election games) and show that if a set-theoretic game $g$ possesses a strictly mixed strategy equilibrium then the safety level value for a player in that game equals the expected payoff it obtains in that equilibrium. Following this, we define the concept of $C$-competitive safety strategies. Roughly speaking, a strategy will be called a $C$-competitive safety strategy, if it guarantees an expected payoff that is $\frac{1}{C}$ of the expected payoff obtained in a Nash equilibrium. We show that in an extended decentralized load balancing setting a 9/8-competitive strategy exists, when the number of players is large. We also discuss extensions of this result to more general settings. In particular, we deal with the cases of arbitrary number of communication lines, and arbitrary different speeds of communication. We show that a ratio of 4/3 can be obtained when we allow arbitrary speeds in two communication lines connecting source to target. We also consider the notion of a $k$-regular network, where $k$ is the ratio between the average communication speed and the lowest speed of communication (in a given set of communication lines), and show that a $k$-competitive safety strategy exists for general $k$-regular networks. Then, we discuss $C$-competitive strategies in the context of Bayesian games. In particular we show the existence of an $e$-competitive safety strategy for a classical first-price auctions setup.

Imagine an agent designed to deal with the communication of a user with different targets. Selecting routes for messages in a multi-agent system is a non-trivial task. The efficiency of the agent depends on the actions selected by other users (and their agents) that try also to communicate with similar targets. In such cases, game-theoretic analysis can identify the Nash equilibria that may emerge in that setting. However, adopting the strategy prescribed by a Nash equilibrium may be quite dangerous for our agent. Other agents may fail to choose strategies prescribed by that equilibrium, and as a result the outcome of our agent can be quite poor. It would have been much better if the agent could have guaranteed similar payoff (to the one obtained in a Nash equilibrium) without relying on other agents' behavior. In computational settings, where (machine and other) failures are possible, and rationality assumptions about participants' behavior should be minimized, a safety-level strategy has a special appeal, especially when it yields a value that is close to the expected payoff obtained in a Nash equilibrium.

Previous work has been concerned with comparing the payoffs that can be obtained by an optimal centralized (and Pareto-efficient) controller to the expected payoffs obtained in the Nash-equilibria of the corresponding game (Koutsoupias & Papadimitriou, 1999).[3] That work is in the spirit of competitive analysis, a central topic in theoretical computer science (Borodin & El-Yaniv, 1998). Our work can be considered as suggesting a complementary approach, comparing the safety-level value to the agent's expected payoff in a Nash equilibrium.

The rest of this paper is organized as follows. In Section 2 we provide some basic definitions and notations. In sections 3 and 4 we deal with simple variants of the load balancing and the leader election problems. We use these as examples for showing that safety-level strategies can be quite competitive and attractive, leading to the value of a Nash equilibrium. This is generalized in section 5 to the context of general $2 \times 2$ games. A discussion of another extension dealing with set-theoretic games is discussed in section 6. In section 7 we deal with several settings of decentralized load balancing, with increasing level of complexity. In particular we show the existence of desired competitive safety strategies for settings with many agents and many possible routes. Section 8 illustrates the use of competitive safety analysis in games with incomplete information.

## 2. Basic Definitions and Notations

A *game* is a tuple $G = \langle N = \{1, \ldots, n\}, \{S_i\}_{i=1}^n, \{U_i\}_{i=1}^n \rangle$, where $N$ is a set of $n$ players, $S_i$ is a finite set of pure strategies available to player $i$, and $U_i : \Pi_{i=1}^n S_i \to \Re$ is the payoff function of player $i$. Given $S_i$, we denote the set of probability distributions over the elements of $S_i$ by $\Delta(S_i)$. An element $t \in \Delta(S_i)$ is called a *mixed strategy* of player $i$. It is called a pure strategy if it assigns probability 1 to an element of $S_i$, and it is called a strictly mixed strategy if it assigns a positive probability to each element in $S_i$. A tuple $t = (t_1, \ldots, t_n) \in \Pi_{i=1}^n \Delta(S_i)$ is called a *strategy profile*. We denote by $U_i(t)$ the expected payoff of player $i$ given the strategy profile $t$. A strategy profile $t = (t_1, \ldots, t_n)$ is a *Nash equilibrium* if $\forall i \in N$, $U_i(t) \geq U_i(t_1, t_2, \ldots, t_{i-1}, t'_i, t_{i+1}, \ldots, t_n)$ for every $t'_i \in S_i$. The Nash equilibrium $t = (t_1, \ldots, t_n)$ is called a *pure strategy Nash equilibrium* if $t_i$ is a pure strategy for every $i \in N$. The Nash equilibrium $t = (t_1, \ldots, t_n)$ is called a *strictly mixed strategy Nash equilibrium* if for every $i \in N$ we have that $t_i$ is a strictly mixed strategy. Given a game $g$ and a mixed strategy of player $i$, $t \in \Delta(S_i)$, the safety level value obtained by $i$ when choosing $t$ in the game $g$, denoted by $val(t, i, g)$, is the minimal expected payoff that player $i$ may obtain when employing $t$ against arbitrary strategy profiles of the other players. A strategy $t'$ of player $i$ for which $val(., i, g)$ is maximal is called a *safely-level strategy* (or a probabilistic maximin strategy) of player $i$. Hence, a safety-level strategy for agent $i$, $s_{safe} \in \Delta(S_i)$ satisfies that

$$s_{safe} \in argmax_{s \in \Delta(S_i)} min_{(s_1, s_2, \ldots, s_{i-1}, s_{i+1}, \ldots, s_n) \in \Pi_{j \neq i} S_j} U_i(s_1, s_2, \ldots, s_{i-1}, s, s_{i+1}, \ldots, s_n)$$

---

3. This work has been extended in e.g. (Roughgarden, 2001; Roughgarden & Tardos, 2002).

A strategy $e \in S_i$ *dominates* a strategy $f \in S_i$ if for every $(s_1, s_2, \ldots, s_{i-1}, s_{i+1}, \ldots, s_n) \in \Pi_{j \neq i} \Delta(S_j)$ we have $U_i(s_1, \ldots, s_{j-1}, e, s_{j+1}, \ldots, s_n) \geq U_i(s_1, \ldots, s_{j-1}, f, s_{j+1}, \ldots, s_n)$, with a strict inequality for at least one such tuple. A game is called *non-reducible* if there do not exist $e, f \in S_i$, for some $i \in N$, such that $e$ dominates $f$. A game is called *generic* if for every $i \in N$, pair of strategies $e, f \in S_i$, and $(s_1, s_2, \ldots, s_{i-1}, s_{i+1}, \ldots, s_n) \in \Pi_{j \neq i} S_j$, we have that $U_i(s_1, \ldots, s_{i-1}, e, s_{i+1}, \ldots, s_n) = U_i(s_1, \ldots, s_{i-1}, f, s_{i+1}, \ldots, s_n)$ only if $e$ and $f$ coincide. In a generic game different strategies of player $i$, assuming a fixed strategy profile for the rest of the players, should lead to different payoffs. This property simply says that in a fixed environment (captured by a strategy profile of the rest of the players), different strategies of player $i$ should lead to somewhat different payoffs (e.g. as a result of their costs, outcomes, etc.) A game is called a $2 \times 2$ game if $n = 2$ and $|S_1| = |S_2| = 2$.

## 3. Decentralized Load Balancing

In this section we consider decentralized load balancing, where two rational players need to submit messages in a simple communication network: a network of two parallel communication lines $e_1, e_2$ connecting nodes $s$ and $t$. Each player has a message that he needs to deliver from $s$ to $t$, and he needs to decide on the route to be taken. The communication line $e_1$ is a faster one, and therefore the value of transmitting a single message along $e_1$ is $X > 0$ while the value of transmitting a single message along $e_2$ is $\alpha X$ for some $0.5 < \alpha < 1$.[4] Each player needs to decide on the communication line to be used for sending its message from $s$ to $t$. If both players choose the same communication line then the value for each one of them drops in a factor of two (a player will obtain $\frac{X}{2}$ if both players choose $e_1$, and a player will obtain $\frac{\alpha X}{2}$ if both players choose $e_2$). In a matrix form, this game can be presented as follows:

$$M = \begin{pmatrix} X/2, X/2 & X, \alpha X \\ \alpha X, X & \alpha X/2, \alpha X/2 \end{pmatrix}$$

**Proposition 1** *The optimal safety-level value for a player in the decentralized load balancing game equals its expected payoff in the strictly mixed strategy equilibrium of that game.*

**Proof:** Consider the following equations for the probability to choose $e_1$ in a symmetric equilibrium, where each player selects $e_1$ with probability $p$ and $e_2$ with probability $1 - p$. This equation is derived from the fact that in a Nash equilibrium every strategy in the support should lead to identical expected payoffs. Notice that by solving this equation we will also prove the existence of a strictly mixed strategy Nash equilibrium.

$$p\frac{X}{2} + (1-p)X = p\alpha X + (1-p)\alpha \frac{X}{2}$$

Hence, $p\frac{X}{2} + X - pX = p\alpha X + \alpha\frac{X}{2} - p\alpha\frac{X}{2}$, and $X - \alpha\frac{X}{2} = p\alpha\frac{X}{2} + p\frac{X}{2}$. This implies that

$$p = \frac{2-\alpha}{1+\alpha}$$

---

4. Notice that here and later in the paper, $X$ is a constant. The important factor is the ratio between the payoffs.

Notice that $0 < p < 1$ as required. The safety level mixed strategy satisfies the following equation. This equation is derived from the fact that the expected payoff of a (mixed) safety-level strategy should be identical for any strategy of the other player.

$$p\frac{X}{2} + (1-p)\alpha X = pX + (1-p)\alpha\frac{X}{2}$$

Hence, $p\frac{X}{2} + \alpha X - p\alpha X = pX + \alpha\frac{X}{2} - \alpha p\frac{X}{2}$. This implies that $pX - \alpha p\frac{X}{2} + p\alpha X - p\frac{X}{2} = \frac{\alpha X}{2}$, and therefore that $p(\frac{X+\alpha X}{2}) = \frac{\alpha X}{2}$. We get:

$$p = \frac{\alpha}{1+\alpha}$$

Notice that the above Nash equilibrium is different from the safety level strategy. However, consider the expected payoff obtained by the Nash equilibrium and by the safety level strategy: The Nash value is:

$$\frac{2-\alpha}{1+\alpha}\frac{X}{2} + \frac{2\alpha-1}{1+\alpha}X$$

The safety level value is:

$$\frac{\alpha}{1+\alpha}\frac{X}{2} + \frac{1}{1+\alpha}\alpha X$$

We will show that these values coincide. It is enough to show that:

$$\frac{2-\alpha}{2(1+\alpha)} + \frac{2\alpha-1}{1+\alpha} = 1.5\frac{\alpha}{1+\alpha}$$

The above however trivially holds since both sides equal $1.5\frac{\alpha}{1+\alpha}$ □

Notice that the above proposition shows that an agent can *guarantee* itself an expected payoff that equals its payoff in a Nash equilibrium of the decentralized load balancing game. This is obtained using a strategy that differs from the agent's strategies in the Nash equilibria of that game (which do not provide that guarantee). Notice that if the players could have used a mediator/correlation devise, and play the game repeatedly, then the mediator could have directed them to the use of strategies leading to a payoff that is higher than the one guaranteed by the safety-level strategy. The use of such mediator/correlation devise, as well as the discussion of repeated games, is beyond the scope of this paper.

## 4. Leader Election: Decentralized Voting

In a leader election setting, the players vote about the identity of the player who will take the lead on a particular task. A failure to obtain agreement about the leader is a bad output, and can be modelled as leading to a 0 payoff. Assume that the players' strategies are either "vote for 1" or "vote for 2", denoted by $a_1, a_2$ respectively, then $U_i(a_j, a_k) > 0$, where $i, j, k \in \{1, 2\}$, and $j = k$. Notice that this setting captures various forms of leader election, e.g. when a player prefers to be selected, when it prefers the other player to be selected, etc. In a matrix form, this game can be presented as follows (where $a, b, c, d > 0$):

$$M = \begin{pmatrix} a, b & 0, 0 \\ 0, 0 & c, d \end{pmatrix}$$

**Proposition 2** *The optimal safety-level value for a player in the leader election game equals its expected payoff in the strictly mixed strategy equilibrium of that game.*

**Proof:** In a strictly mixed Nash equilibrium we have that the probability $q$ of choosing $a_1$ by player 2 should satisfy:

$$qU_1(a_1, a_1) = (1-q)U_1(a_2, a_2)$$

The above equality is implied by the fact that any pure strategy in the support of the mixed strategy for an agent, in a Nash equilibrium, should yield the same expected payoff (otherwise, deviation will be rational.) Hence, the above equality captures the fact that the strategy of player 2 in equilibrium should be selected in a way that the utility for agent 1 when using either $a_1$ or $a_2$ will be the same.

Similarly, the probability $p$ of choosing $a_1$ by player 1 should satisfy

$$pU_2(a_1, a_1) = (1-p)U_2(a_2, a_2)$$

Hence, a strictly mixed strategy Nash equilibrium exists, where $q = \frac{U_1(a_2,a_2)}{U_1(a_1,a_1)+U_1(a_2,a_2)}$ and $p = \frac{U_2(a_2,a_2)}{U_2(a_1,a_1)+U_2(a_2,a_2)}$. As can be seen from the above equations a strictly mixed strategy equilibrium exists. Consider now w.l.o.g player 1. The expected payoff it obtains in the above equilibrium is $qU_1(a_1, a_1) = \frac{U_1(a_1,a_1)U_1(a_2,a_2)}{U_1(a_1,a_1)+U_1(a_2,a_2)}$. Player 1's safety level strategy satisfies the following, where $p'$ is the probability of choosing $a_1$:

$$p'U_1(a_1, a_1) = (1-p')U_1(a_2, a_2)$$

Hence, $p' = \frac{U_1(a_2,a_2)}{U_1(a_1,a_1)+U_1(a_2,a_2)}$. Notice that $p' = q$. The safety level value will be therefore: $p'U_1(a_1, a_1) = \frac{U_1(a_1,a_1)U_1(a_2,a_2)}{U_1(a_1,a_1)+U_1(a_2,a_2)}$. We get that the Nash equilibrium and safety level strategies are different, but their expected payoffs for the players coincide. □

Notice that the above proposition shows that a agent can *guarantee* itself an expected payoff that equals its payoff in a Nash equilibrium of the leader election game.[5] As in the decentralized load balancing game, this is obtained using a strategy that differs from the agent's strategies in the Nash equilibria of that game (which do not provide that guarantee).

## 5. Safety Level in General $2 \times 2$ Games

The results presented in the previous sections refer to 2-person 2-choice variants of central problems occurring in computational contexts. Given the encouraging results in the framework of these basic settings, we wish to consider two types of extensions:

1. Generalize the results to a broader family of simple games.

2. Generalize the results to more general CS-related settings, dealing in particular with games with many players, as found in load-balancing settings.

---

5. The reader should not confuse the fact that $p' = q$ with similarity between safety-level and Nash equilibrium. Indeed, $p'$ refers to the probability of choosing $a_1$ by player 1, while $q$ refers to the probability of choosing that action by player 2.

In this section we deal with the first point. Later, and in particular in section 7, we will deal with the second one. It is of interest to see whether our results in sections 3-4 can be extended to other forms of $2 \times 2$ games. Notice that the load balancing and the leader election settings can be represented as non-reducible generic $2 \times 2$ games. The same is true with regard to the game presented by Aumann:

$$M = \begin{pmatrix} 2,6 & 4,2 \\ 6,0 & 0,4 \end{pmatrix}$$

Non-reducible generic games are an attractive concept. Having dominated strategies in the game do not add to the understanding of the interaction, since these strategies can be safely ignored. The fact a game is generic is also quite appealing: it is quite natural to assume that a pair of actions should lead to different outcomes when we fix the rest of the environment. We can show:

**Theorem 1** *Let $G$ be a $2 \times 2$ non-reducible generic game. Assume that the optimal safety level value of a player is obtained by a strictly mixed strategy, then this value coincides with the expected payoff of that player in a Nash equilibrium of $G$.*

**Proof:** Denote the strategies available to the players by $a_1, a_2$. Use the following notation: $a = U_1(a_1, a_1), b = U_1(a_1, a_2), c = U_1(a_2, a_1), d = U_1(a_2, a_2), e = U_2(a_1, a_1), f = U_2(a_1, a_2), g = U_2(a_2, a_1), h = U_2(a_2, a_2)$

In a matrix form, the above will be presented as:

$$M = \begin{pmatrix} a,e & b,f \\ c,g & d,h \end{pmatrix}$$

If a strictly mixed strategy Nash equilibrium exists then it should satisfy that:

$$qa + (1-q)b = qc + (1-q)d$$

and

$$pe + (1-p)g = pf + (1-p)h$$

where $p$ and $q$ are the probabilities for choosing $a_1$ by players 1 and 2, respectively. We get that we should have $qa + b - qb = qc + d - qd$, which implies that $q(a - b - c + d) = d - b$. Similarly, we get that we should have $pe + g - pg = pf + h - ph$, which implies that $p(e - g - f + h) = h - g$. Hence, in a strictly mixed strategy Nash equilibrium we should have:

$$q = \frac{d-b}{a-b-c+d}$$

and

$$p = \frac{h-g}{e-g-f+h}$$

Notice that since the game is generic then $d \neq b$. If $d > b$ then if $q$ is not strictly in between 0 and 1 then $c > a$ which will contradict non-reducibility. If $d < b$ then in if $q$ is not strictly in between 0 and 1 then $a > c$, which also contradicts non-reducibility. Similarly, since the game is generic then $h \neq g$. If $h > g$ then if $p$ is not strictly in between 0 and 1 then $f > e$ which will contradict non-reducibility. If $h < g$ then in if $p$ is not strictly in between 0 and 1 then $e < f$, which also contradicts non-reducibility. Given the above we get that $p$ and $q$ define a strictly mixed strategy equilibrium of $G$. Consider now the safety level strategy of player 1. If player 1 chooses $a_1$ with probability $p'$ then it satisfies that:

$$p'a + (1-p')c = p'b + (1-p')d$$

This implies that we need to have $p'a + c - p'c = p'b + d - p'd$, which implies $p'(a-c-b+d) = d-c$. Hence, we have

$$p' = \frac{d-c}{a-c-b+d}$$

and

$$1 - p' = \frac{a-b}{a-c-b+d}$$

Compute now the expected payoff for player 1 in the strictly mixed Nash equilibrium, given that $1 - q = \frac{a-c}{a-b-c+d}$, we have that:

$$qa + (1-q)b = \frac{(d-b)a + (a-c)b}{a-b-c+d} = \frac{da - cb}{a-b-c+d}$$

The expected payoff of the safety level strategy for player 1 will be:

$$p'a + (1-p')c = \frac{(d-c)a + (a-b)c}{a-b-c+d} = \frac{da - cb}{a-b-c+d}$$

Hence, we get that the expected payoffs of the Nash equilibrium and the safety level strategies for player 1 coincide. The computation for player 2 is similar. □

### 5.1 The Case of Pure Safety-Level Strategies

The reader may wonder whether the previous result can be also proved for the case where there are no restrictions on the structure of the safety-level strategy of the game $g$. In several AI contexts, the discussion is on pure maximin strategies, where probabilistic behavior is not considered. Of course, probabilistic maximin strategies are more powerful, and in many cases the best safety level is obtained only by a mixed strategy and not by a pure one. However, it will be of interest to consider the case where the safety-level strategy is a pure one. As we now show, there exists a generic non-reducible $2 \times 2$ game $g$, where the optimal safety level strategy for a player is pure, and the expected payoff for that player is lower than the expected payoff for that player in all Nash equilibria of $g$. Consider a game $g$, where $U_1(1,1) = 100, U_1(1,2) = 40, U_1(2,1) = 60, U_1(2,2) = 50$, and $U_2(1,1) = 100, U_2(1,2) = 210, U_2(2,1) = 200, U_2(2,2) = 90$. In a matrix form this game looks as follows:

$$M = \begin{pmatrix} 100, 100 & 40, 210 \\ 60, 200 & 50, 90 \end{pmatrix}$$

It is easy to check that $g$ is generic and non-reducible. In particular, there are no dominated strategies, and the payoffs obtained by each player for different strategy profiles are different from one another. The game has no pure Nash equilibria. In a strictly mixed strategy equilibrium the probability $q$ of choosing $a_1$ by player 2 should satisfy $100q + 40(1 - q) = 60q+50(1-q)$, i.e. that $60q+40 = 10q+50$, $q = 0.2$. In that equilibrium the probability that player 1 will choose $a_1$ is $p = 0.5$, and the expected payoff of player 1 is $100q+40(1-q) = 52$. The safety-level strategy for player 1 is to perform $a_2$, guaranteeing a payoff of 50, given that $(a_2, a_2)$ is a saddle point in a zero-sum game where the payoffs of player 2 are taken to be the complement to 0 of player 1's original payoffs. Hence, the value of the safety level strategy for player 1 is $50 < 52$. □

## 6. Beyond $2 \times 2$ Games

The leader election game is an instance of a more general set of games: *set-theoretic games*. In a set theoretic game the sets of strategies available to the players are identical, and the payoff of each player is uniquely determined by the *set* of strategies selected by each player. For example, in a 2-person set-theoretic game we will have that $U_1(s,t) = U_1(t,s), U_2(s,t) = U_2(t,s)$ for every $s, t \in S_1 = S_2$. Notice that set-theoretic games are very typical to voting contexts. In a typical voting context we care about the votes, but not about the indentity of the voters. We can prove the following:

**Proposition 3** *Given a 2-person set theoretic game $g$ with a strictly mixed strategy Nash equilibrium, then the value of an optimal safety level strategy of a player equals its expected payoff in that equilibrium.*

**Proof:** Let $S = S_1 = S_2 = \{s_1, s_2, \ldots, s_l\}$. Let $t = (t_1, t_2)$ be a strictly mixed strategy Nash equilibrium. Denote the tuple of probabilities associated with $t_i$ by $(p_{i_1}, \ldots, p_{i_l})$ ($i \in \{1,2\}, |S_1| = |S_2| = l$). In a strictly mixed Nash equilibrium we have that the expected payoff of player 1 is:

$$\Sigma_{j=1}^{l} p_{2_j} U_1(s_e, s_j) \quad (*)$$

for every $1 \leq e \leq l$. Consider now a strategy $f$ of player 1 that assigns probability $p_{2_j}$ to strategy $s_j$. Then, for every strategy $s_e$ selected by player 2, the expected payoff of $f$ is given by

$$\Sigma_{j=1}^{l} p_{2_j} U_1(s_j, s_e) = \Sigma_{j=1}^{l} p_{2_j} U_1(s_e, s_j) = (*)$$

This implies that the safety level strategy for player 1 yields an expected payoff that is identical to the expected payoff for player 1 in the above equilibrium. Similar reasoning can be applied for player 2. □

## 7. Competitive Safety Strategies

Let $S$ be a set of strategies. Consider a family of games $(g_1, g_2, \ldots, g_j, \ldots)$ where $i$ is a player at each of them, its set of strategies at each of these games is $S$, and there are $j$ players, in addition to $i$, in $g_j$. As an example, consider a family of decentralized load balancing settings. The $(n-1)$-th game in this extended load-balancing setting will consist of $n$ players, one of them is $i$. The players submit their messages along $e_1$ and $e_2$. The payoff for player $i$ when participating in an $n$-person decentralized load balancing game is $\frac{X}{k}$ (resp. $\frac{\alpha X}{k}$) if he has chosen $e_1$ (resp. $e_2$) and additional $k - 1$ participants have chosen that communication line. A mixed strategy $t \in \Delta(S)$ will be called a *C-competitive safety strategy* if there exists some constant $C > 0$, such that

$$\lim_{j \to \infty} \frac{nash(i, g_j)}{val(t, i, g_j)} \leq C$$

where $nash(i, g_j)$ is the lowest expected payoff player $i$ might obtain in some equilibrium of $g_j$, and $val(t, i, g_j)$ is the expected payoff guaranteed for $i$ by choosing $t$ in the game $g_j$. The extended decentralized load balancing setting [6] is a typical and basic network problem. If $C$ is small, a $C$-competitive safety strategy for that context will provide a useful protocol of behavior. We can show:

**Theorem 2** *There exists a 9/8-competitive safety strategy for the extended decentralized load-balancing setting.*

**Proof:** Consider the following strategy profile for the players in an $n$-person decentralized load balancing game: players $\{1, 2, \ldots, \lceil \frac{1}{1+\alpha} n \rceil\}$ will choose $e_1$, and the rest will choose $e_2$. W.l.o.g we assume that $i = 1$ is the player for which we will make the computation of expected payoffs. It is easy to verify that the above strategy profile is an equilibrium of the game, with an expected payoff for player $i$ that is bounded above by

$$\frac{X(1+\alpha)}{n} \quad (**)$$

Intuitively, this equilibrium is obtained by partitioning the players in a way where the payoff for using the communication lines are (almost) equal. Consider now the following strategy $t$ for player $i$: select $e_1$ with probability $\frac{\alpha}{1+\alpha}$ and select $e_2$ with probability $\frac{1}{1+\alpha}$. Notice that $t$ (if adopted by all participants) is not a Nash equilibrium. However, we will show that it is a competitive safety strategy for small $C > 0$. Consider an arbitrary number of participants $n$, where $\beta(n-1)$ of the other (i.e. excluding player $i$) $n-1$ participants use $e_2$ while the rest use $e_1$, for some arbitrary $0 \leq \beta \leq 1$. The expected payoff obtained using $t$ will be:

$$\frac{1}{1+\alpha} \frac{\alpha X}{\beta(n-1)+1} + \frac{\alpha}{1+\alpha} \frac{X}{(1-\beta)(n-1)+1}$$

This value is greater or equal to:

$$\frac{1}{1+\alpha} \frac{\alpha X}{\beta n + 1} + \frac{\alpha}{1+\alpha} \frac{X}{(1-\beta)n + 1}$$

The above equals

$$\frac{X\alpha}{1+\alpha} \left[ \frac{1}{\beta n + 1} + \frac{1}{(1-\beta)n + 1} \right]$$

Simplifying the above we get:

$$\frac{X\alpha}{1+\alpha} \frac{n+2}{(1+\beta n)(n - \beta n + 1)} \quad (***)$$

---

6. Here and later the term *extended* load-balancing setting refers to a family of games as above.

Dividing (**) by (***) we get that the ratio is:

$$\frac{(1+\alpha)^2}{\alpha} \frac{(\beta-\beta^2)n^2 + n + 1}{n(n+2)}$$

When $n$ approaches infinity the above ratio approaches

$$\frac{(1+\alpha)^2}{\alpha}(\beta-\beta^2)$$

Given that $0.5 \leq \alpha < 1$ and $0 \leq \beta \leq 1$ we get that the above ratio is bounded by $9/8$ as desired. $\square$

### 7.1 Extensions: Arbitrary Speeds and $m$ Links

In this section we generalize the result obtained in the context of decentralized load balancing to the case where we have $m$ parallel communication lines leading from source to target. The value obtained by the agent (w.l.o.g. agent 1) when submitting its message along line $i$, where $n_i$ agents have decided to submit their messages through that line is given by $\frac{X \cdot \alpha_i}{n_i}$, where $1 = \alpha_1 \geq \alpha_2 \geq \cdots \geq \alpha_m > 0$. Our extension enables us to handle the general binary case where $0 < \alpha < 1$, as well as to discuss cases where a safety level strategy can be very effective in the general $m$-lines situation. Using the ideas developed for the case $m=2$, we can now show:

**Theorem 3** *There exists a $\frac{\Sigma_{i=1}^m \alpha_i \Sigma_{i=1}^m \Pi_{j \neq i}\alpha_j}{m^2 \Pi_{i=1}^m \alpha_j}$–competitive safety strategy for the extended decentralized load-balancing setting, when we allow $m$ (rather than only 2) parallel communication lines, and arbitrary $\alpha_i$'s.*

**Proof:** Following the ideas of the previous theorem, there exists an equilibrium where agent 1 obtains at most $(X/n)\Sigma_{i=1}^m \alpha_i$. Intuitively, in this equilibrium the players are distributed in a way where the payoff for using the different communication lines are (almost) identical. In particular, agents $\{1, 2, \ldots, \lceil \frac{\alpha_1}{\Sigma_{i=1}^m \alpha_i} n \rceil\}$, where $\alpha_1 = 1$ will be assigned to communication line 1, and hence agent 1's payoff will be as prescribed.

Consider the following strategy for each of the agents: choose communication line $i$ with probability

$$\frac{\Pi_{j \neq i}\alpha_j}{\Sigma_{i=1}^m \Pi_{j \neq i}\alpha_j}$$

Given the above, the expected payoff of agent $i$ can be minimized (using similar ideas to the ones in the proof of Theorem 2), by splitting the other agents equally among the communication lines. Hence, the expected payoff of the agent is at least:

$$\Sigma_{i=1}^m \frac{\alpha_i X \Pi_{j\neq i}\alpha_j}{(1+\frac{n}{m})\Sigma_{i=1}^m \Pi_{j\neq i}\alpha_j} = \frac{m^2 X}{m+n}\frac{\Pi_{j=1}^m \alpha_j}{\Sigma_{i=1}^m \Pi_{j\neq i}\alpha_j}$$

Hence, the ratio between the expected payoff in the Nash equilibrium and the expected payoff that can be guaranteed is bounded by:

$$\frac{m+n}{m^2 n}(\Sigma_{i=1}^m \alpha_i)\frac{\Sigma_{i=1}^m \Pi_{j\neq i}\alpha_j}{\Pi_{j=1}^m \alpha_j}$$

373

The above implies, when $n$ is large the existence of an $\frac{\Sigma_{i=1}^m \alpha_i \Sigma_{i=1}^m \Pi_{j \neq i} \alpha_j}{m^2 \Pi_{j=1}^m \alpha_j}$–competitive strategy. $\square$

In the general binary case, where $\alpha_1 = 1$, and $\alpha_2 = \alpha$, where $0 < \alpha \leq 1$, the above implies the existence of an

$$\frac{(1+\alpha)^2}{4\alpha}$$

competitive strategy.

**Corollary 1** *Given an extended load balancing setting, where $m = 2$, with arbitrary speeds of the communication lines ($0 < \alpha \leq 1$), there exists a $\frac{4}{3}$-competitive strategy.*

**Proof:** To see the above notice that $1 + \alpha < \frac{(1+\alpha)^2}{4\alpha}$ if and only if $\alpha < 1/3$ and that $\frac{(1+\alpha)^2}{4\alpha}$ is decreasing in the interval $(0, 1]$. Hence, by considering a strategy prescribed by the above theorem when $\alpha \geq 1/3$ and selecting $e_1$ otherwise, we are guaranteed a ratio of at most $1 + 1/3 = 4/3$. $\square$

Consider now the general $m$-links (i.e. $m$ parallel communication lines) case. The average network quality (or speed), $Q$, can be defined as $\frac{\Sigma_{i=1}^m \alpha_i}{m}$. A network will be called $k$-regular if $\frac{Q}{\alpha_m} \leq k$. Many networks are $k$-regular for small $k$. For example, if $\alpha_m \geq 0.5$ as before, then the network is 2-regular regardless of the number of edges.

**Corollary 2** *Given a $k$-regular network, there exists a $k$-competitive safety strategy for the extended decentralized load-balancing setting, when we allow $m$ (rather than only 2) parallel edges.*

**Proof:** To show the above, observe that

$$\frac{\Sigma_{i=1}^m \alpha_i \Sigma_{i=1}^m \Pi_{j \neq i} \alpha_j}{m^2 \Pi_{j=1}^m \alpha_j} = \frac{Q}{m} \frac{\Sigma_{i=1}^m \Pi_{j \neq i} \alpha_j}{\Pi_{j=1}^m \alpha_j}$$

The latter is smaller or equal to

$$\frac{Q}{m} \frac{m}{\alpha_m} = k$$

as desired. $\square$

Together, Theorem 3 and corollaries 1 and 2 extend the results on decentralized load balancing to the general case of $m$ parallel communication lines.

## 8. Competitive Safety Analysis in Bayesian Games

The results presented in the previous sections refer to games with complete information. The games we have studied in this context refer to fundamental settings in the AI and game theory intersection, and deal with issues such as congestion. In this section, we show that our ideas can be applied to games with incomplete information as well.

In a game with incomplete information the payoff for a player given the behavior of the set of players is private information of that player. In order to illustrate competitive safety analysis in games with incomplete information, we have chosen to consider a very basic mechanism, the first-price auction. The selection of first-price auction is not an accident.

Auctions are fundamental to the theory of economic mechanism design[7], and among the auctions that do not possess a dominant strategy, assuming the independent private value model, first-price auctions are probably the most common ones.

We consider a setting where a good $g$ is put for sale, and there are $n$ potential buyers. Each such buyer has a valuation (i.e. maximal willingness to pay) for $g$ that is drawn from a uniform distribution on the interval of real numbers $[0, 1]$. This valuation is the private information that the agent has. The exact valuation is known only to the agent, while the distribution on agent valuations are commonly known. The valuations are assumed to be independent from one another. In a first price auction, each potential buyer is asked to submit a bid for the good $g$. We assume that the bids of a buyer with valuation $v$ is a number in the interval $[0, v]$.[8] The good will be allocated to the bidder who submitted the highest bid (with a lottery to determine the winner in a case of a tie). The auction setup can be defined using a Bayesian game.[9] In this game the players are the potential bidders, and the payoff of a player with valuation $v$ is $v-p$ if he wins the good and pays $p$, and 0 if he does not get the good. As the reader can see, the distinguished feature of such games is that the player's utility function depends on the agent's private valuation, and therefore it is known only to it. The equilibrium concept can be also extended to the context of Bayesian games. In the auction setup an agent's strategy is a function from its valuations to monetary bids. A strategy profile will be in equilibrium if an agent's strategy is the best response against the other agents' strategies given the distribution on these agents' valuations. In particular, in equilibrium of the above game the bid of a player with valuation $v$ is $(1 - \frac{1}{n})v$.

Given the above, the expected payoff of an agent with valuation $v$, will be $\frac{v^n}{n}$. As before, the question is whether we can guarantee a payoff that is proportional to the expected payoff in equilibrium.

Before discussing an appropriate strategy, we should emphasize a formal issue with regard to competitive safety strategies in Bayesian games. Notice that in our definition of competitive safety strategies, we assume that the player's competitive action should be independent of the number of players. On the other hand, as suggested by the equilibrium analysis above, behavior in first-price auction may heavily depend on the number of players. In order to address this issue, we make use of the revelation principle, discussed in the economic mechanism design literature. The revelation principle tells us that one can replace the above-mentioned first-price auction with the following auction: each bidder will be asked to reveal his valuation, and the good will be sold to the bidder who reported the highest valuation; if agent $i$ who reported valuation $v'$ will turn out to be the winner then he will be asked to pay $(1 - \frac{1}{n})v'$. In this mechanism a player will submit bids in between 0 and $\frac{n}{n-1}v$. It turns out that reporting the true valuation is an equilibrium of that auction, and that it will yield (in equilibrium) the same allocation, payments, and expected utility to the participants, as the original auction. It is convenient to consider the above *revelation mechanism*, since when facing *any* number of participants, a bidder's strategy in equilibrium will always be the same.

---

7. For a general discussion of mechanism design see (Mas-Colell, Whinston, & Green, 1995), Chapter 23, and (Fudenberg & Tirole, 1991), Chapter 7).
8. In general, buyers may submit bids that are higher than their valuations, but these strategies are dominated by other strategies, and their existence will not effect the equilibrium discussed in this paper.
9. A formal definition and exposition of Bayesian games can be found in (Fudenberg & Tirole, 1991).

Given the above, a first-price auction setup will be identified with a family of (Bayesian) games $(g_1, g_2, \ldots)$ where $g_j$ is the Bayesian game associated with (the revelation mechanism of) first-price auction with $j+1$ potential buyers. The definition of $C$-competitive strategies can now be applied to the above context as well.

**Theorem 4** *There exists an e-competitive strategy for the first-price auction setup.*

**Proof:** When player 1 with valuation $v$ submits the bid $b$ in an auction with additional $n-1$ players, its worst case payoff is

$$\int_{v_2=0}^{\frac{n-1}{n}b} dv_2 \int_{v_3=0}^{\frac{n-1}{n}b} dv_2 \cdots \int_{v_n=0}^{\frac{n-1}{n}b} (v - \frac{n-1}{n}b) dv_n$$

The above says that in order to win, player $i$'s bid should be higher than the other players' bids. Each player's bid in the revelation mechanism is however at most $\frac{n}{n-1}$ times its valuation, and therefore we should integrate over valuations that are at most $\frac{n-1}{n}$ times player $i$'s bid. If agent $i$ will be the winner then he will gain $v - \frac{n-1}{n}b$ when he bids $b$ and his valuation is $v$ (given the rules of the revelation mechanism). The above is maximized when

$$\frac{d}{db}(v - \frac{n-1}{n}b)(\frac{n-1}{n}b)^{n-1} = 0$$

Hence, the expected value is maximized when $(n-1)vb^{n-2} = \frac{n-1}{n}nb^{n-1}$, i.e. when $b = v$. We therefore get that the safety-level strategy coincides in this case with the equilibrium strategy. The expected payoff in equilibrium can be shown to be $v^n/n$. The expected payoff guaranteed by the above strategy will be

$$(\frac{n-1}{n}v)^{n-1} \frac{1}{n}v$$

The ratio between the safety level value and the equilibrium value is therefore bounded above by $(\frac{n-1}{n})^n$, which is greater or equal to $\frac{1}{e}$, and approaches it when the number of players approaches infinity. □

An interesting observation is that in the above theorem, the safety-level strategy is identical to the equilibrium strategy. This connection occurs although the game is not a 0-sum game. It is interesting to observe that since we consider revelation mechanisms then the safety-level strategy turns out to be independent of the number of participants. Our result can also be obtained if we consider standard first-price auctions, rather than the revelation mechanisms associated with them; nevertheless, this will require us to allow a player to choose its action knowing the number of potential bidders (as in the corresponding equilibrium analysis).

## 9. Discussion

Some previous work in AI has attempted to show the potential power of decision-theoretic approaches that do not rely on classical game-theoretic analysis. In particular, work in theoretical computer science on competitive analysis has been extended to deal with rationality constraints (Tennenholtz, 2001), in order to become applicable to multi-agent systems. We introduced competitive safety analysis, bridging the gap between the normative AI/CS approach and classical equilibrium analysis. We have shown that the observation, due to Aumann, that safety-level strategies may yield the value of a Nash equilibrium in games that are not zero-sum, provides a powerful normative tool for computer scientists and AI researchers interested in protocols for non-cooperative environments. We have illustrated the use and power of competitive safety analysis in various contexts. We have shown general results about $2 \times 2$ games, as well as about games with many participants, and introduced the use of competitive safety analysis in the context of decentralized load balancing, leader election, and auctions. Notice that our work is concerned with a normative approach to decision making in multi-agent systems. We make no claims as for the applicability of this approach for descriptive purposes, i.e. for the prediction of how people will behave in the corresponding situations. Although there exists much literature on the failure of Nash equilibrium, it is still the most powerful concept for action prediction in multi-agent systems. The setting of decentralized load balancing discussed as part of this paper is central to game theory and its applications.[10] Given the importance of this setting from a CS perspective, providing robust agent protocols for that setting is a major challenge to work in multi-agent systems. In order however to build robust protocols, relying on standard equilibrium analysis might not be satisfactory, and safety guarantees are required. Our work suggests protocols and analysis for providing such guarantees, bridging the gap between classical AI/decision-theoretic reasoning and equilibrium analysis in game theory.

## Acknowledgements

This work has been carried out when the author was on a sabbatical leave with the computer science department at Stanford university. A preliminary version of this paper appears in the proceedings of AAAI-2002.

---

10. See the literature on potential and congestion games, e.g. (Monderer & L.S.Shapley, 1996; Rosenthal, 1973).